\documentclass[draft,showpacs,preprintnumbers,twocolumn]{revtex4}
\usepackage{amsfonts}
\usepackage{amsmath}
\usepackage{amssymb}
\usepackage{graphicx}
\usepackage{dcolumn}
\usepackage{bm}
\usepackage{mathcomp}
\usepackage{subfigure}
\usepackage{lipsum}

\newenvironment{proof}[1][Proof]{\begin{trivlist}
\item[\hskip \labelsep {\bfseries #1}]}{\end{trivlist}}

\newcommand{\qed}{\nobreak \ifvmode \relax \else
      \ifdim\lastskip<1.5em \hskip-\lastskip
      \hskip1.5em plus0em minus0.5em \fi \nobreak
      \vrule height0.75em width0.5em depth0.25em\fi}

\begin{document}

\title{An information theoretical analysis of quantum optimal control}


\author{S. Lloyd$^{1}$, S.~Montangero$^2$}

\affiliation{$^1$Massachusetts Insitiute of Technology, Department of Mechanical Engineering, Cambridge MA 02139 USA, 
\\ $^2$Institut f\"ur Quanteninformationsverarbeitung, Universit\"at Ulm, 89069 Ulm, Germany.}

\date{\today}
\begin{abstract}
We show that if an efficient classical representation of the dynamics exists, 
optimal control problems on many-body quantum systems can be solved efficiently with finite precision. 
We show that the size of the space of parameters necessary to solve quantum 
optimal control problems defined on pure, mixed states and unitaries is polynomially bounded from the size of the of the set 
of reachable states in polynomial time. We provide a bound for the minimal time necessary to perform the  optimal process 
given the bandwidth of the control pulse, that is the continuous version of the Solovay-Kitaev theorem. 
We explore the connection between entanglement present in the system and 
complexity of the control problem, showing that one-dimensional slightly entangled dynamics 
can be efficiently controlled. Finally, we quantify how noise affects the presented results.
\end{abstract}


\maketitle

Quantum optimal control lies at the heart of the modern quantum revolution, as it allows to match the stringent 
requirements needed to develop quantum technologies, to develop novel quantum protocols and 
to improve their performances~\cite{rabitz09}.  
Along with the increased numerical and experimental capabilities developed in recent years, problems of 
increasing complexity have been explored and recently a lot of attention has been devoted to the application 
of optimal control (OC) to many-body quantum dynamics:
OC has been applied to information processing in quantum wires~\cite{qwire}, the crossing 
of quantum phase transitions~\cite{qpt}, the generation of many-body squeezed or entangled states~\cite{ent}, 
chaotic dynamics~\cite{chaos}, unitary transformations~\cite{rabitz12}.
Recent studies have been devoted to the 
understanding of the fundamental limits of OC in terms of energy-time relations 
(time-optimal)~\cite{QSL} and its robustness against perturbations~\cite{kallush11,kosloff13}. 

These exciting developments call for the development of 
a general framework to understand when and under which conditions is it possible 
to solve a given OC problem in a many-body quantum system. 
Indeed, due to the exponential growth of the Hilbert space with the number of constituents, 
solving an OC problem on a many-body system is in general highly inefficient: 
the algorithmic complexity (AC) of exact time-optimal problems can be super-exponential~\cite{rabitz12}. 
However,  limited precision, errors and practical limitations naturally introduce 
a finite precision both in the functional to be minimized and on the 
total time of the transformation. The smoothed complexity (SC)
has been introduced recently to cope with this situation 
to describe the ``practical" complexity of solving a problem in the real world with finite precision.  
It has been shown that the SC can be drastically different from the AC:
indeed the AC --which is defined by the scaling of the worst case-- might be practically 
irrelevant as the worst case might be never found in practice~\cite{blaser12}.
A paradigmatic case is that of the simplex algorithm applied to linear programming problems: 
it is characterized by an exponential AC in the dimension of the searched space, 
however the SC is only polynomial, that is,  the worst case disappears in presence of perturbations~\cite{spielman04}. 

In this letter, we perform an information theoretical analysis providing a first step towards the 
theoretical understanding of the complexity of OC problems in many-body quantum systems. 
We present a counting 
argument to bound the size of the space of parameters needed to solve OC problems 
defined over the set of time-polynomially reachable states. We explore the implications of this result 
in terms of SC identifying some classes of problems that can be 
efficiently solved. We characterize the effects of noise in the control field and of the entanglement 
present during the system dynamics. 
We finally provide an information-time bound, relating the bandwidth of the control field with the minimal 
time necessary to achieve the optimal transformation. 
 
A quantum OC problem can be stated as follows: 
given a dynamical equation 
\begin{equation}
\dot \rho = \mathcal L (\rho, \gamma(t));
\label{liuv}
\end{equation}
with boundary condition $\rho(t=0)= \rho_0$ where $\rho$ is the density matrix describing 
a quantum system defined on an Hilbert space $\mathcal{H} = \mathbb{C}^N$, and $\mathcal L $
the Liouvillian operator with the unitary part generated by an Hamiltonian 
\begin{equation}
H= H_D + \gamma(t) H_C,
\label{ham}
\end{equation}
where $\gamma(t)$ is a time-dependent control field, and $H_D$ and $H_C$ the drift and control Hamiltonian respectively. 
For simplicity here we consider the case where only a single control field is present (the generalization is straightforward) and
we work in adimensional units. From now on we focus on finite-size Hilbert space of dimension $N$, as any quantum system with limited energy and limited in space is effectively finite-dimensional.  
Eq.~\eqref{liuv} generates a set of states depending on the control field $\gamma(t)$ and on the initial state $\rho_0$:  
the manifold that is generated for every $\gamma (t)$ defines the set  of reachable states $\mathcal W$  
with dimension $D_{\mathcal W}(N)$~\cite{schirmer03}. 
If the system is controllable --i.e. the operators $H_D, H_C$ generate the complete dynamical Lie algebra-- 
the manifold ${\mathcal W}$ is the complete space of density matrix operators and its dimension is    
$D_{\mathcal W} = N^2$ for an $N-$dimensional Hilbert space, where for $n$ d-level quantum systems $N=d^n$. 
Given a goal state $\bar \rho$ the problem to be solved is to find a control pulse 
$\bar \gamma(t)$ that drives the system from a reference state $\rho_0$ 
within an $\epsilon$-ball around the goal state $ \bar \rho$.
Equivalently, the OC problem can be 
expressed as a functional minimization of the form 
\begin{equation}
\min_{\gamma(t)} \mathcal F (\rho_0, \bar \rho,\gamma(t), [\lambda_\imath]),
\label{problem}
\end{equation}
where the functional $\mathcal F$ might also include constraints introduced 
via Lagrange multipliers $\lambda_\imath$. 
The functional $\mathcal F$ is minimised  by an (not necessary unique) optimal 
$\bar \gamma(t)$, that identifies a final state $\rho_f$ such that $ || \rho_f - \bar \rho|| < \epsilon$ in some norm $|| \cdot ||$. 

We now recall the definition of the information content of the control pulse $\gamma(t)$ as we 
show in the following that it is intimately related to the complexity of the OC problem.
The information  (number of bits $b_\gamma$) carried by the control pulse 
$\gamma(t)$ is given by the classical channel capacity $C$ times the pulse duration $T$. 
In the simple case of a noiseless channel, the channel capacity is given by Hartley's law, thus
\begin{equation}
b_\gamma =  T \,  \Delta\Omega \, \kappa_s
\label{bg}
\end{equation}
where $ \Delta\Omega$ is the bandwidth, and $\kappa_s = \log(1 + \Delta \gamma/\delta \gamma) $ is the bit depth of 
the control pulse $\gamma(t)$, and $\Delta \gamma = \gamma_{max} -\gamma_{min}$ and $\delta \gamma$ are the maximal and minimal allowed variation of the field~\cite{chitode}.
Note that given an uniform sampling rate of the signal $\delta t $, $ T \,  \Delta\Omega = T/\delta t  = n_s $ where $n_s$ 
is the number of sampling points of the signal.  
Any optimization method of choice 
depends on these $n_s$ variables, i.e. $n_s$ defines the dimension of the input of the optimisation problem.   
We thus define the dimension of the quantum OC problem $\mathcal{D}$ as follows: 
{Given a dynamical law of the form of Eq.\eqref{liuv},  a reference initial state $\rho_0$ and any possible goal state in the set 
reachable states $\mathcal W$, the dimension of the quantum OC problem is defined by the minimal number of 
independent degrees of freedom $\mathcal{D}$ in the OC field
 necessary to achieve the desired transformation up to precision 
$\varepsilon$.}
Notice that $\mathcal{D}$ might be the minimal number of sampling points $n_s$, of independent bang-bang controls, 
of frequencies present in the control field or the dimension of the subspace of functions the control field has non-zero projection on. 

From now on we consider the physical situations where the control is performed in some finite time $t  \in [0,T]$, 
with bounded control field and bounded Hamiltonians, e.g. $||H_D|| =||H_C ||=1$ and $ \gamma(t) \in [\gamma_{min}:\gamma_{max}] \, \forall t$. 
The aforementioned physical constraints, naturally introduce a new class of states, that we define as follows:
{The set of time-polynomial reachable states $\mathcal W^+ \subseteq \mathcal W$ 
is the set of states that can be reached  (with finite energy) with precision $\varepsilon$ 
in polynomial time as a function of the set size $ D_{\mathcal W^+}(N) \le D_{\mathcal W}(N) $.}
This is the class of interesting states from the point of view of OC, as if a state can 
be reached only in exponential time there is no need of OC at all: 
in exponential time any reachable state is reached also with a constant Hamiltonian. 
Similarly to standard definitions, we define a time-polynomial reachable system 
if all states can be reached (with precision $\varepsilon$) 
in polynomial time by means of at least one path (i.e.  $D_{\mathcal W^+}=D_{\mathcal W}$) 
and a time-polynomially controllable system if $\mathcal W^+$ is equal to the whole Hilbert space.  
Notice that if the bound on the strength of the control $\gamma_{max}$ is relaxed we have  
$D_{\mathcal W^+}=D_{\mathcal W}$.
Given the above definitions, we can state the following:
~\\~\\
{\bf Theorem} The size $\mathcal{D}$ of a quantum OC problem in $\mathcal W^+$ 
up to precision $\varepsilon$ is a polynomial function of the size of the manifold of the time-polynomial reachable states $D_{\mathcal W^+}$. 

\begin{proof}
We first prove that the dimension of the problem 
is bounded from below by $D_{\mathcal W^+}$ and then that is bounded from above by a polynomial function of 
$D_{\mathcal W^+}$. \newline
{Lower bound}: We divide the complete set of time-polynomial reachable states $\mathcal{W^+}$ in balls 
of size $\varepsilon^{D_{\mathcal W^+}}$. The number of $\varepsilon$-balls necessary to cover the whole set 
$\mathcal{W^+}$ is $\varepsilon^{- D_{\mathcal W^+}}$ and one of them 
identifies the set of states that live around the state  $\bar \rho$ within a radius $\varepsilon$.  
The information content of the OC field 
must be at least sufficient to specify the $\varepsilon$-ball surrounding the goal state, that is $b_\gamma  \ge b_S^-$, where 
$b_S^- = \log \varepsilon^{-D_{\mathcal W^+}}$.
Finally one obtains
\begin{equation}
\varepsilon \ge  2^{-\frac{T \,  \Delta\Omega \, \kappa_S}{D_{\mathcal W^+}}}.
\label{error}
\end{equation}
Setting a maximal precision (e.g. machine precision) expressed in bits $\kappa_\varepsilon = - \log_2 \, \varepsilon$ 
results in $n_s  \kappa_s /D_{\mathcal W^+} = \kappa_\varepsilon$; and imposing $\kappa_\varepsilon = \kappa_s$ we obtain 
\begin{equation}
n_s  \ge D_{\mathcal W^+}.
\end{equation}
{Upper bound}:
The goal state belongs to the set of time-polynomial states $\bar \rho \in \mathcal W^+$, thus a path of finite length $L$ that connects the initial and goal 
states in polynomial time exists. The maximum of (non-redundant) information that provides the solution to 
the problem is the information needed to describe the complete path $b_S^+$. Setting the desired precision $\varepsilon$, this is equal to 
$\log \varepsilon^{-D_{\mathcal W^+}}$ bit of information for each $\varepsilon$-ball needed to cover the path times the number of balls
$n_\varepsilon$.  The latter is given by 
\begin{equation}
n_\varepsilon = L / \varepsilon \le T v_{max} / \varepsilon= Poly(D_{\mathcal W^+}) v_{max}/ \varepsilon
\end{equation}
where $L$ is the length of the path, $v_{max}$ is the maximal allowed velocity along the path due to the bounded energy. 
In conclusion, we obtain that 
\begin{equation}
b_S^+ = \frac{Poly(D_{\mathcal W^+}) v_{max}}{\varepsilon} \log \varepsilon^{-D_{\mathcal W^+}},
\label{bs+}
\end{equation}
that implies together with the condition $b_\gamma  \leq b_S^+$
\begin{equation}
 Poly'(D_{\mathcal W^+}) v_{max}/{\varepsilon}  \ge n_s  
\end{equation}
As $n_s$ is bounded by a polynomial function of $D_{\mathcal W^+}$, thus $\mathcal{D}=Poly(D_{\mathcal W^+})$ $\blacksquare$
\end{proof}

Notice that the lower bound holds in general for any reachable state in $\mathcal W$ and can be saturated, as recently shown in~\cite{caneva13}. 
On the other hand, the upper bound diverges for $\varepsilon \to 0$,  as finding the exact solution of the control problem might be as difficult as super exponential~\cite{rabitz12}. 
The theorem has a number of interesting practical and theoretical implications that we present in the rest of the paper.   

{\it Complexity -} 
The aforementioned theorem poses the basis to set the SC of solving the OC problem. 
An algorithm recently introduced to solve complex quantum OC problems, the Chopped RAndom Basis (CRAB) optimisation, 
builds on the fact that the space of the control pulse $\bar \gamma(t)$ 
is limited from the very beginning to some (small) value $\mathcal{D}$, and then solves the problem by means of a direct 
search method as the simplex algorithm. Recently, numerical evidence has been presented that this algorithm efficiently founds 
exponentially precise solutions as soon as $\mathcal{D} \ge D_\mathcal{W}$~\cite{CRAB}. 
This result can be put now on solid ground as
under fairly general conditions OC problems are equivalent to linear programming~\cite{taksar96} 
and linear programming can be solved via simplex algorithm with polynomial SC~\cite{blaser12}: 
thus, the CRAB optimisation solves with polynomial SC OC problems with dimension $\mathcal{D}$. 
More formally, one can make the following statement: 
The class of  OC problems that satisfy the hypothesis (H1-H3) of Ref.~\cite{taksar96}, is characterised by
a polynomial SC in the size of the problem ${\mathcal D}$.
In conclusion, studying the scaling of the dimension of the control problem $\mathcal{D} = Poly(D_\mathcal{W^+})$ is of fundamental 
interest to understand and classify our capability of efficiently control quantum systems. The first results in this direction can be obtained observing the influence of  
the integrability of the quantum system on $D_\mathcal{W^+}$, resulting in the following properties: 

1 - The size $\mathcal{D}$ of a generic OC problem defined on time-polynomial controllable 
non-integrable $n$-body quantum system is exponential with the number of constituents $n$. 
Indeed the dynamics of a controllable non integrable many-body quantum system explores the whole Hilbert space, 
i.e. the set of time-polynomial reachable states is the whole Hilbert space, that is $D_{\mathcal W^+} = N^2$ 
($D_{\mathcal W^+}= N$ for pure states). 

On the contrary, despite the exponential growth of the Hilbert space, the size of $\mathcal W^+$ for integrable systems 
is at most linear in the number $n$ of constituents of the system, that implies together with the theorem above that:

2 - The size $\mathcal{D}$ of OC problems defined on time-polynomially controllable integrable many-body quantum system, is polynomial with $n = \log_d(N)$. 
Notice that this statement generalizes a theorem that has been proven for the particular case of tridiagonal Hamiltonian systems presented in~\cite{fu01}. 

Finally, there exists a class of intermediate dynamics that despite in principle might explore an exponentially big Hilbert space, are confined in a 
corner of it and can thus be efficiently represented. The simplest example of this class of problems is mean-field dynamics, however more 
generally, to this class of dynamics belongs for example 
those that can be represented efficiently by means of a tensor-network as t-DMRG~\cite{schollwock}. 
We can thus state the following:

3 - The dimension $\mathcal{D}$  of an OC problem defined on a dynamical process that can be described efficiently by a tensor network, 
e.g. in one dimension a matrix product state, is polynomial in the number of system components $n$. 
The dimension of the set of the time-polynomial reachable states $\mathcal {W^+}$ that can be efficiently represented by a tensor network 
scales as  $D_{\mathcal W^+} \leq D_{\mathcal W} \leq Poly(n) \cdot T$ where $T$  is the total time of the evolution and  $Poly(n)$ is the 
dimension of the biggest tensor network state represented during the time evolution. 
Notice that, although the previous statement is in principle valid in all dimensions, it has practical implications mostly in one-dimensional systems as 
much less efficient representations of the dynamics are known in dimensions bigger than one~\cite{2DTN}. 

We can now link directly the entanglement present in the system during its dynamics with the complexity of controlling it:

4 - Time evolution of slightly entangled one-dimensional many-body quantum systems can be efficiently represented via Matrix Product States with $D_{\mathcal {W^+}} \leq D_{\mathcal {W}} = O (T \, d \, 2^{2 S} n) $ parameters, where $S$ is the maximal Von Neumann entropy of any bipartition present in the system. Thus, systems with $S \propto \log (n) $ for every time  can be efficiently controlled. 

We stress that the entanglement present in the system is not uniquely correlated with the complexity of the OC problem: 
indeed due to the previous results, integrable systems (also highly entangled) are efficiently controllable, as shown recently in~\cite{caneva13}. 
On the contrary, as said before, highly entangled dynamics of non integrable systems, for which it does 
not exists an efficient representation as $S \propto n $ are exponentially difficult to control. 
In conclusion, the size of the control problem depends on the dimension of the manifold over which the dynamics 
takes place. This can be simply understood by considering the scenario where the dynamics over which the 
control problem is defined is restricted to the space of two eigenstates of a complex many-body hamiltonian, 
each of them highly entangled w.r.t some local bases. If one has access to a direct coupling between them, the complexity 
of the OC problem is not more than that of a simple Landau-zener process (independently from the entanglement 
present in the system) as the manifold is effectively two-dimensional. 
However, this is not generally the case, as one has usually access to some local (or global) operator, and the dynamic of the system is 
not in general restricted to two states. In the case of non integrable systems, a generic couple of initial and goal states projects
on exponentially many basis states independently of the chosen basis, while for integrable states it exists a base where the states have 
a simple representation. Thus, the minimal amount of information needed to solve the quantum OC problem is exponential and polynomial respectively. In between, there is the class of TN-efficiently representable dynamics, for which we know how to build an efficient 
representation and correspondingly we know how to efficiently solve the OC problem. 

{\it Time bounds -} 
Manipulating  Eq.~\eqref{error} applied to the whole set of reachable states $\mathcal W$ 
we achieve a bound for the minimal time needed to achieve the desired transformation as a function of the control bandwidth: 
The minimal time needed to reach a given final state in $D_{\mathcal W}$ with precision $\varepsilon$ at finite bandwidth is
\begin{equation}
T \ge \frac{D_{\mathcal W}}{\Delta\Omega \, \kappa_S} \log(1/\varepsilon)
\end{equation}
or again, under the assumption that  $\kappa_\varepsilon = \kappa_s$:
\begin{equation}
T \ge \frac{D_{\mathcal W}}{\Delta\Omega }.
\label{SK}
\end{equation}
The previous relation is a continuous version of the Solovay-Kitaev theorem: it provides an estimate of the minimal time needed to perform an optimal process given a finite band-width. 
Notice also that the bandwidth provides the average bits rate per second, thus this results coincides with the intuitive expectation that the minimal time needed to perform an optimal quantum process 
is the time necessary to ``inform" the system about the goal state given that the control field has only a finite bit transmission rate. 

We recall that there is a time-energy bound, known as quantum speed limit that in its general form is 
\begin{equation}
T_{QSL} \ge \frac{d(\rho_0, \rho_G)}{ \overline \Lambda},
\end{equation}
where $d(\cdot,\cdot)$ is the distance and $\overline{\Lambda} = \int_0^T || \mathcal{L} ||_p dt / T$ with $|| \cdot ||_p$ the p-norm~\cite{QSL}.
The best efficient process saturates both bounds, that implies  
$\Delta \Omega \propto D_{\mathcal W}$;
thus the bandwidth of the time-optimal pulse in general should scale as the dimension of the space $\mathcal W$, 
requiring exponential higher frequencies for non integrable many-body quantum systems and 
thus practically preventing its physical realization. 

{\it Noise -} In presence of noise Eq.~\eqref{bg} has to be modified: in the following we consider a common scenario however this analysis can be adapted to the specific noise considered. 
For gaussian white noise, according to Shannon-Hartley theorem the channel capacity is 
$k_s = \log (1 + S/N)$, 
where $S/N$ is the signal to noise power ratio~\cite{chitode}. Thus, following the same steps as before we obtain that 
\begin{equation}
\varepsilon \ge  (1+S/N)^{-\frac{n_s}{D_{\mathcal W}}}, 
\label{errornoise}
\end{equation}
and similarly 
\begin{equation}
T \ge \frac{D_{\mathcal W}}{\Delta\Omega }  \frac{\log (1/\varepsilon)}{\log (1+ S/N)}.
\label{SKnoise}
\end{equation}
For small noise to signal ratio ($N/S \ll 1$) the previous bound results in 
$\varepsilon \gtrsim (N/S)^{n_s/D_{\mathcal W}}$ which together with the fact that $n_s$ has to be a polynomial function 
of $D_{\mathcal W^+}$ show that the control problem is in general exponential sensitive to the problem dimension. 
However, if one saturates the lower bound on the complexity of the optimal field, i.e. $n_s = D_{\mathcal W}$, 
the sensitivity to Gaussian white noise become linear in the noise to signal ratio. That is, the effects of the noise on 
the optimal transformation are negligible if the noise level is below the error, $N/S \lesssim \varepsilon$.
As requiring the optimal transformation to be more precise than the error on the control signal is somehow unnatural, 
this relation demonstrate that OC transformations are in general robust with respect to noise, as recently 
observed~\cite{montangero05}. At the same time, for $\varepsilon \lesssim N/S$ this results agrees with the scaling 
for exact optimal transformations recently found in~\cite{kosloff13}.


{\it Control of unitaries -} The aforementioned statements also hold for the generation of unitaries as the differential equation governing the evolution of the time evolution 
operator
$\imath \hbar \dot U(t) = H(t) U(t)$
is formally equivalent to Eq.~\eqref{liuv} replacing the density matrix with the time evolution operator $U(t)$, 
the reference state with the identity operator, and the goal state with the unitary to be generated. 



{\it Observability - } As any controllable system is also observable by a coherent controller~\cite{lloyd00}, the previous definitions and results can be  
straightforward applied to the complexity of observing a many-body quantum system with precision $\varepsilon$. 

In conclusion, we have shown that if one allows a finite error (both in the goal state and in time) 
as it typically occurs in any practical application of OC, 
what can be efficiently simulated can also be optimally controlled and that the optimal 
solution is in general robust with respect to perturbation on the control field. 
Notice that the presented results are valid both for open and closed loop OC. 

We thank T. Calarco, A. Negretti, and P. Rebentrost for discussions and feedback. 
SM acknowledge support from the DFG via SFB/TRR21 and from the EU projects SIQS and DIADEMS.



\end{document}